# N-opcode Analysis for Android Malware Classification and Categorization

BooJoong Kang, Suleiman Y. Yerima, Kieran McLaughlin, Sakir Sezer
Centre for Secure Information Technologies (CSIT)
Queen's University Belfast
Belfast, Northern Ireland, United Kingdom
{b.kang, s.yerima, kieran.mclaughlin, s.sezer}@qub.ac.uk

*Abstract*—Malware detection is a growing problem particularly on the Android mobile platform due to its increasing popularity and accessibility to numerous third party app markets. This has also been made worse by the increasingly sophisticated detection avoidance techniques employed by emerging malware families. This calls for more effective techniques for detection and classification of Android malware. Hence, in this paper we present an *n*-opcode analysis based approach that utilizes machine learning to classify and categorize Android malware. This approach enables automated feature discovery that eliminates the need for applying expert or domain knowledge to define the needed features. Our experiments on 2520 samples that were performed using up to 10-gram opcode features showed that an f-measure of 98% is achievable using this approach.

## I. INTRODUCTION

The proliferation of malware on the mobile platform is an increasing problem especially on Android, due to its growing popularity and availability of millions of apps from the official Google play store, third-party app markets and various online sources. The volume of new applications appearing frequently is too large for manual examination of each application for malicious behavior to be feasible. Hence, this process does not scale very easily to large numbers of applications. Previous studies have also shown that traditional signature-based approaches, which most antivirus scanners employ, fails to be effective at detecting new malware due to their increasing adoption of sophisticated detection avoidance techniques and the need for frequent update of signature databases.

Android malware detecting is currently an active area of research. Consequently, there is a growing volume of work on automated detection that utilize machine learning techniques. Various methods have been proposed based on examining the dynamic application behavior [1, 2, 3], requested permissions [4, 5, 6], API calls [6, 7, 8, 22] etc. However these methods are often still largely reliant on expert analysis to design or determine the discriminative features that are passed to the machine learning system used to make the final classification decision.

Some recent studies such as [9, 10, 11] have investigated the effectiveness of *n*-gram opcodes (or *n*-opcodes as referred to in this paper) extracted from the disassembled application byte-code as a means for Android malware detection. The advantage of the use of an opcode based technique is the ability to automatically learn features from raw data directly rather than specifying them beforehand through expert analysis. For example, in previous works such as [8] where the discriminative features are based on API calls, expert analysis provides the selection of the 'most interesting' features (i.e. API calls methods, signatures, etc.). Unlike the opcode based techniques, this could limit the scope of the application of machine learning algorithms by excluding potentially useful learning information.

Hence, in this paper we investigate *n*-opcode analysis for Android malware detection using machine learning on real datasets. We study this approach and analyze its efficacy for both application classification to detect malware and also malware categorization (i.e. classification into known families). Unlike previous works that experimented with opcodes of up to 5-grams only, the work presented in this paper analyzed up to 10-grams sequences whilst also considering both their frequencies and binary counts. The rest of the paper is organized as follows: Section II reviews related work; Section III presents the *n*-opcode analysis technique; the evaluation experiments and discussions are presented in Section IV while Section V concludes the paper.

## II. RELATED WORK

In this section we review related work on malware detection. The two main approaches commonly applied to malware detection are static analysis and dynamic analysis. Static analysis involves disassembling the program binary in order to extract features, while dynamic analysis involves running the program in an emulator or instrumented hardware in order to extract characteristic actions performed by the program. Static analysis has the advantage of being faster and may enable greater code coverage than dynamic analysis, while dynamic analysis may be less prone to code obfuscation. Some previous works such as [14], [15], and [16], have combined the two approaches.

Learning based approaches using hand-designed features have been applied extensively to both dynamic [1-3], [18-20]

and static [6, 7, 8, 22] malware detection. For example, the authors of [7] studied a static analysis approach to Android malware detection based on 179 features manually derived from API calls, intents, permissions and commands that were combined with ensemble learning. Their approach was evaluated on a dataset of 2925 malware apps and 3938 benign apps. A variety of similar approaches to static malware detection have used similarly derived features, but with different classifiers such as support vector machine (*SVM*) [12], Naïve Bayes [17], and *k*-nearest neighbor [6]. Malware detection approaches have also been proposed that use static features derived exclusively from the permissions requested by the application [4, 5].

In contrast with approaches using high-level hand-designed features, such as relevant permissions or API calls, *n*-grams based malware detection uses sequences of low-level opcodes as features. The *n*-grams features can be used to train a classifier to distinguish between malware and benign software [9], or to classify malware into different families [10]. Perhaps surprisingly, even a 1-gram based feature, which is simply a histogram of the number of times each opcode is used, can distinguish malware from benign software [13]. The length of the *n*-gram used [9] and number of *n*-gram sequences used in classification [13] can both have an effect on the accuracy of the classifier.

In this paper, an *n*-gram approach is also investigated. However, unlike the previous papers we develop an approach that enables the use of longer *n*-grams thus analyzing up to 10-grams whereas the previously reported works utilized up to 5-grams [9, 11]. Also, [10] and [13] only considered 1-gram opcode features. Furthermore, unlike [9] which was only based on binary information, we investigate both frequency and binary information thus allowing for greater information coverage. Finally, we provide in-depth analysis that provides clear insight into why our approach achieve high classification accuracy with various machine learning algorithms.

III. N-OPCODE ANALYSIS

In this section, we explain how to extract *n*-opcodes from Android applications and how to select *n*-opcodes that will enable optimal malware classification and categorization. The following two subsections describe the process of the *n*-opcode extraction and the feature selection and also include statistical results from our dataset.

*A. N-opcode Extraction*

The *n*-opcode extraction of an application consists of disassembling the application and extracting *n*-opcodes. An Android application can be delivered as a compressed file, an Android application package (*apk*) file, containing a manifest file, resource files and *Dalvik* executable (*dex*) files. The *dex* files contain the application source code and can be disassembled using *baksmali* [26]. As a result of disassembling, *baksmali* generates a set of *smali* files for the *dex* file, where each *smali* file represents a single class that contains all the methods of the class. Each method contains human-readable *Dalvik* bytecode (shortly instructions) and each instruction consists of a single opcode and multiple operands. We discard the operands and only extract *n*-opcodes from each method. The resulting output of the *n*-opcode extraction is a vector of the unique *n*-opcodes from all the classes of the application. The vector contains the frequency of each unique *n*-opcode. The overview of the *n*-opcode extraction is shown in Fig. 1.

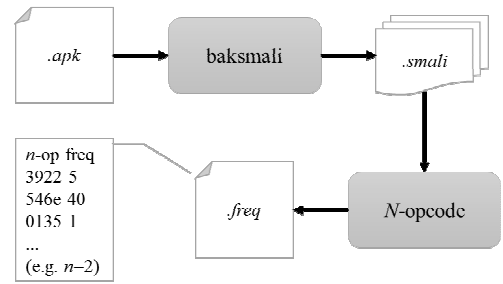

Fig. 1. *N*-opcode extraction

Any *n*-gram based method faces the prospect of exponential increase in the number of unique *n*-grams as *n* is increased. It is expected that a similar trend will be observed with *n*-opcodes as well. Fig. 2 and TABLE I. show the number of unique *n*-opcodes for different *n* from our dataset, which consists of 1260 samples of malware and 1260 benign samples. In the malware classification (*MC*) study, we processed the *n*-opcode extraction on all the 2520 samples and counted the number of unique *n*-opcodes for different *n*, with *n* ranging from 1 to 10.

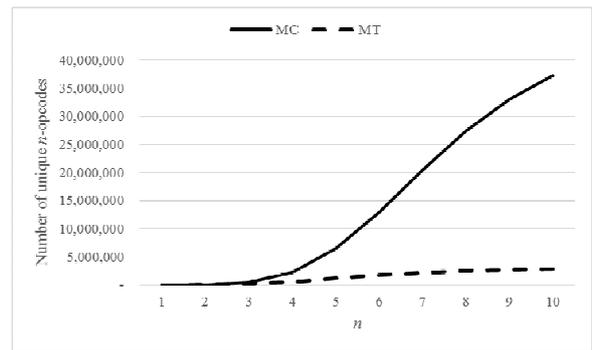

Fig. 2. Number of unique *n*-opcodes vs. *n*

TABLE I. NUMBER OF UNIQUE *N*-OPCODES FOR DIFFERENT VALUES OF *N*

| *n* | *MC* | *MT* |
|---|---|---|
| 1 | 214 | 210 |
| 2 | 22,371 | 14,550 |
| 3 | 399,598 | 154,483 |
| 4 | 2,201,377 | 557,526 |
| 5 | 6,458,246 | 1,145,025 |
| 6 | 12,969,857 | 1,724,771 |
| 7 | 20,404,473 | 2,177,621 |
| 8 | 27,366,890 | 2,491,721 |
| 9 | 33,024,116 | 2,695,226 |

| n | MC | MT |
|---|---|---|
| 10 | 37,186,183 | 2,823,729 |

It can be observed that there is no exponential increase in the number of unique $n$-opcodes. Instead, it increases linearly at first and gradually becomes stable. One explanation for this is that some methods may contain less instructions than $n$ therefore those $n$-opcodes will not appear in such methods. Another reason is that a bigger $n$ is likely to generate a smaller number of $n$-opcodes than a smaller $n$ generates from the same method. For example, a method with 7 instructions has 6 pieces of 2-opcodes, 5 pieces of 3-opcodes, 4 pieces of 4-copdoes and so on. Therefore, the maximum number of the unique $n$-opcodes for a method is in inverse proportion to $n$. In the malware categorization (*MT*) study, the number of unique $n$-opcodes also becomes stable despite utilizing only malware samples for the study. Even though there is no exponential increase, nonetheless an excessive number of unique $n$-opcodes that could cause a huge overhead in further processes results. The number of the unique 10-opcodes in *MC* and *MT* were observed to be 37186183 (about 37M) and 2823729 (about 2.8M), respectively.

## B. Feature Selection

Since the number of unique $n$-opcodes is excessive, it is difficult to run machine learning algorithms on the original data. One of solutions for this issue is feature selection i.e. a process of identifying the best features and is a widely used approach to filter out less important features. In the feature selection stage, we measure the information gain of each feature and subsequently filter out the less important features that have the lowest information gains.

Entropy is a measure of the uncertainty of a random variable. The entropy of a variable $X$ is defined in (1) below and the entropy of $X$ after observing values of a variable $Y$ is defined in (2), where $P(x_i)$ is the prior probabilities for all values of $X$ and $P(x_i|y_i)$ is the posterior probabilities of $X$ given the values of $Y$.

$$H(X) = -\sum P(x_i)\log_2(P(x_i)) \quad (1)$$

$$H(X/Y) = -\sum P(y_j) \sum P(x_i/y_j)\log_2(P(x_i/y_j)) \quad (2)$$

The amount by which the entropy of $X$ decreases reflects additional information about $X$ provided by $Y$ and is called information gain [21], given by (3).

$$IG(X/Y) = H(X) - H(X/Y) \quad (3)$$

According to this measure, a feature $Y$ is regarded as a better indicator than a feature $Z$ for a class $X$, if $IG(X|Y) > IG(X|Z)$. We rank the features by the information gain and select the high ranked features.

In order to compute the *IG*, we used an implementation of the information gain in *WEKA* [24]. However, the program could not handle the large data input from the $n$-opcode feature files and frequently encountered an out of memory error (on a Linux PC with 16GB RAM). So in order to overcome this problem, we segmented the data into several smaller chunks (in multiple *.arff* files) and computed the *IG*s on smaller data. This worked because the information gain algorithm computes the score of a feature independently. Hence, we processed the information gain on each small set of features and merge the results together at the end as illustrated in Fig. 3.

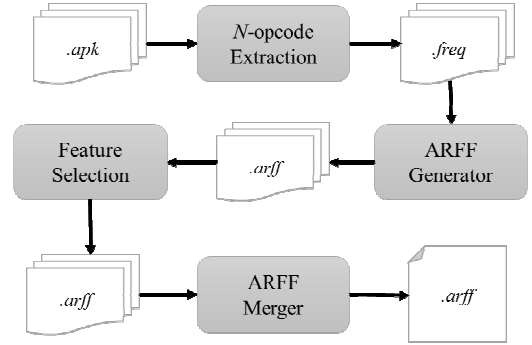

Fig. 3. The overview of our analysis process

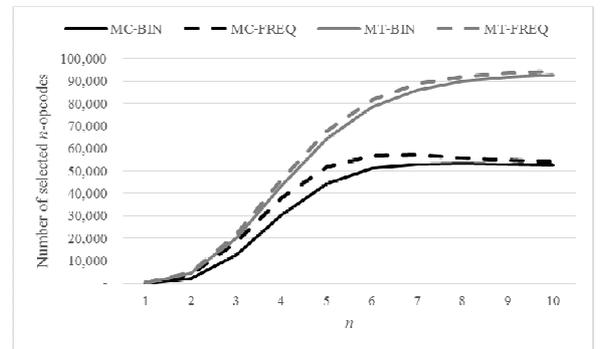

Fig. 4. Number of selected $n$-opcodes

TABLE II. NUMBER OF SELECTED $N$-OPCODES

| n | MC | | MT | |
|---|---|---|---|---|
| | Binary | Frequency | Binary | Frequency |
| 1 | 29 | 191 | 164 | 199 |
| 2 | 2,121 | 4,155 | 4,567 | 4,973 |
| 3 | 12,808 | 18,411 | 20,376 | 21,851 |
| 4 | 30,293 | 37,992 | 43,538 | 46,253 |
| 5 | 44,532 | 51,845 | 64,450 | 67,837 |
| 6 | 51,213 | 56,850 | 78,453 | 81,736 |
| 7 | 53,079 | 57,086 | 86,146 | 88,839 |
| 8 | 53,139 | 55,856 | 90,024 | 92,076 |
| 9 | 52,857 | 54,837 | 91,966 | 93,600 |
| 10 | 52,588 | 54,080 | 92,878 | 94,239 |

Fig. 4 and TABLE II. show the number of selected $n$-opcodes with information gain greater than 0.1 for different $n$. As the table illustrates, the number of selected $n$-opcodes increased as $n$ becomes larger. This means that we gain more information by increasing *n*. However, the increase in the number of selected *n*-opcodes reaches a saturation point as *n* increases. Because of this observation, we expect that impact of the increase in *n* on classification accuracy to peak at this saturation point where further increase in *n* will have little effect. Another interesting observation is that the number of selected *n*-opcodes for frequency is slightly greater than the number of selected *n*-opcodes for binary. Note that frequency and binary refer to the number of counts of the *n*-opcode. For the former it is the overall count, while for the latter it is 1 or 0 denoting presence and absence respectively.

IV. EVALUATION

In this section, we evaluate the performance of *n*-opcodes for malware classification and categorization with different *n*. Evaluation and detail analyses for malware classification and categorization are presented in the following subsections, respectively. In each subsection, we also compare the performance of two different data types: binary and frequency.

Our dataset consists of malware from the Android Malware Genome project [25] and has a total of 2520 applications, of which 1260 are benign and 1260 are malware from 49 different malware families. Labels are provided for the malware family of each sample. The benign samples were collected from the Google play store and have been checked using *VirusTotal* [23] to ascertain that they were highly probable to be malware free. We use four different machine learning algorithms: Naïve Bayes (*NB*), support vector machine (*SVM*), partial decision tree (*PART*) and random forest (*RF*) and utilize *WEKA* as the framework. The following experimental results are reported using the weighted average f-measure, which is based on the precision and recall, over 10-fold cross validation.

*A. Malware Classification*

In the malware classification study, samples are classified into one of two classes: benign or malware. We evaluated the performance of malware classification with two different data types of *n*-opcodes: binary and frequency.

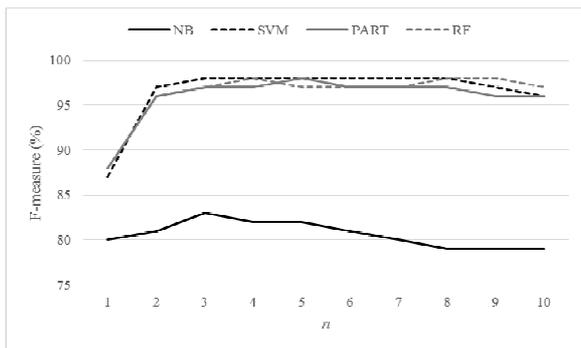

Fig. 5. Malware classificatoin results with binary *n*-opcodes

Fig. 5 shows the results of the binary *n*-opcodes. With the exception of *NB*, the performance of the other three algorithms were similar although *SVM* shows the best performance in most cases. The f-measure increases as *n* is increased but no more increase is observed when *n* is greater than 3. The f-measure even tends to decrease when *n* is greater than 7. This trend is similar with the change in the number of selected *n*-opcodes in Fig. 4. *NB* shows the worst performance but shows the same trend. The best f-measure is 98% and *SVM* performed with the best f-measure when *n* is 3.

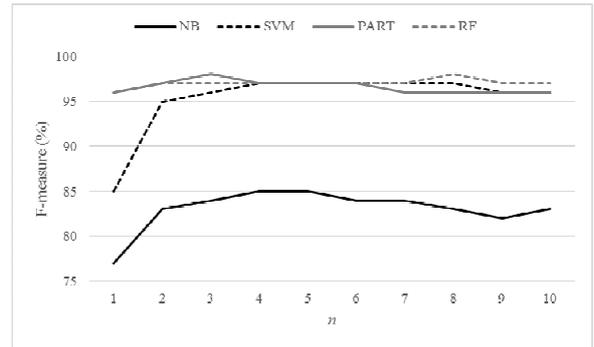

Fig. 6. Malware classificatoin results with frequency *n*-opcodes

As shown in Fig. 6, the frequency *n*-opcodes show similar results with the binary *n*-opcodes. One interesting observation is that the frequency *n*-opcodes show a good performance when *n* is 1 (i.e. no *n*-gram is applied) compared to the binary *n*-opcodes.

TABLE III. TOP TNE *N*-OPCODES FOR MALWARE CLASSIFICATION

| Rank | Our Findings | | | [9] |
|---|---|---|---|---|
| | **3** | **4** | **5** | **5** |
| 1 | **08076e** | 6e0c086e# | 6e0c086e0c# | 136e6e6e0c |
| 2 | 08546e# | 08546e0c# | 2038071f6e# | 1c6e6e0c6e* |
| 3 | 0c086e | 0c086e0c# | 0854380854# | 1d546e0a39 |
| 4 | 220870# | 08546e0a# | 38071f6e0a# | 210135461a |
| 5 | 12086e# | 08540854# | 0854085408# | 0c1a6e0a33 |
| 6 | 390f6e# | 38071f6e# | 0c08546e0c# | 12123c0e22* |
| 7 | 085412# | 390f6e0a# | 0c086e0c6e# | 7154626e28 |
| 8 | 390f54 | 08543808# | 08546e0854# | 6e2820381f |
| 9 | 085208# | **08076e0c** | 20381f2822 | 0d076e289c* |
| 10 | 085438# | 3808546e# | 08546e0c08# | 16313d740e* |

\#: only found in malware samples, \*: only found in malware samples,
bold : mostly found in malware samples, shadow : found in our top ranked *n*-opcodes

TABLE III. shows the top ten binary *n*-opcodes for malware classification. We also present the top ten 5-opcodes of [9] (which were the only ones provided) to compare with our findings. Even though no same 5-opcodes commonly appear between the two top ten 5-opcodes (in the two rightmost columns), we found six of the top ten 5-opcodes from [9], highlighted in the last column, in our selected 5-opcodes as the 11770th, 12597th, 13241th, 13260th, 15927th and 25662th ranked features, respectively. This could be attributed to the fact that our benign samples differed even though we utilized the same malware samples as [9]. An interesting observation is that three of the overlapping *n*-opcodes ('12123c0e22', '0d076e289c' and '16313d740e') are also only found in malware samples just as observed in [9]. Another observation is that our top hundred

5-opcodes are mostly found in benign samples. As we highlighted *n*-opcodes only found in benign samples with '#', it can be seen from the table that there were more of the top ten *n*-opcodes only found in benign samples. This would indicate that the unique *n*-opcodes from benign samples were a strong contributing factor for malware classification.

The last observations is that three 4-opcodes ('08546e0c', '08546e0a' and '3808546e') and three 5-opcodes ('0c08546e0c', '08546e0854' and '08546e0c08') were extensions of the second ranked 3-opcode '08546e'. We also found 51 4-opcodes and 75 5-opcodes from the low ranked *n*-opcodes, which were extensions of the same 3-opcode. From our evaluation, all these extensions only occurred in benign samples. Based on this observation we think it would be interesting to investigate training the machine learning classifiers with combination of different lengths *n*-opcode features in future. In this work, the machine learning classifiers are trained with features of the same length *n*-opcodes.

### B. Malware Categorization

In the malware categorization study, samples are classified into one of existing malware families. We also evaluated the performance of malware categorization with two different data types of *n*-opcodes. Fig. 7 shows the results of binary *n*-opcodes which shows a similar trend with the malware classification results. *SVM* shows the best f-measure of 98%, with its f-measure becomes steady when *n* is 4.

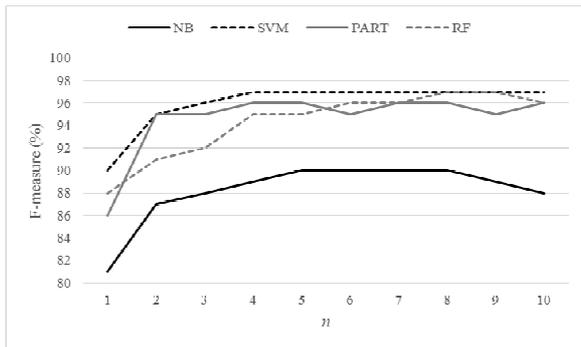

Fig. 7. Malware categorization results with binary *n*-opcodes

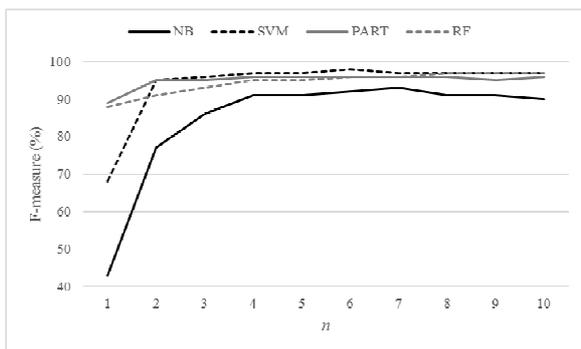

Fig. 8. Malware categorization results with frequency *n*-opcodes

Fig. 8 shows the results of frequency *n*-opcodes for malware categorization. Again, *SVM* shows the best f-measure of 98%, when *n* is 6 but *SVM* and *PART* show a high f-measure of 95%, when *n* reaches only 2. The frequency 1-opcodes do not show a better performance compared to the binary 1-opcodes, as it does in malware classification. In contrast to malware classification, malware categorization is a multi-class classification, which is considered as a more difficult problem compared to the binary classification. TABLE IV. shows the top ten binary *n*-opcodes for malware categorization. Since we have no data from other research for comparison, we only analyze our data for malware categorization. We observed that there is no *n*-opcode that was found only (exclusively) in a single family within the top ten *n*-opcodes.

TABLE IV. TOP TEN *N*-OPCODES FOR MALWARE CATEGORIZATION

| Rank | *n* | | |
|---|---|---|---|
| | 3 | 4 | 5 |
| 1 | 36121a | 700c390e | 22621a701a |
| 2 | 3c0e22 | 621a701a | 621a701a71 |
| 3 | 1a1607 | 123c0e22 | 123c0e2270 |
| 4 | 616174 | 12123c0e | 3c0e22706e |
| 5 | 3b7428 | 313b7428 | 12123c0e22 |
| 6 | 313b74 | 3c0e2270 | 0c6e0c236e |
| 7 | 123c0e | 1a706e15 | 6e0a386e71 |
| 8 | 3d740e | 3922701c | 0c22702271 |
| 9 | 0a8104 | 3d740e0d | 0c1a221a6e |
| 10 | 289c08 | 8104085a | 0c1a706e15 |

In terms of classification accuracy, *SVM* shows the best performance in all the cases. However, the accuracy is not the only factor when choosing a machine learning algorithm. Another factor is the speed performance and TABLE V. shows the time overhead of the machine learning algorithms from the malware classification with the binary *n*-opcodes. *RF* shows the best performance in terms of both training and prediction speeds. Even though *PART* has the highest training overhead it is still considered a suitable classifier because once trained, classification is fast as well.

TABLE V. TIME OVERHEAD OF MACHINE LEARNING ALGORITHMS

| *n* | Machine Learning Algorithms | | | |
|---|---|---|---|---|
| | NB | SVM | PART | RF |
| 1 | 0.02/0.01 | 1/0 | 0.14/0 | 0.14/0 |
| 2 | 0.31/0.22 | 1.53/0.01 | 3.26/0 | 0.27/0 |
| 3 | 2.18/1.5 | 5.65/0.03 | 25.82/0 | 0.34/0.01 |
| 4 | 6.41/3.36 | 12.8/0.1 | 62.38/0.01 | 0.51/0.01 |
| 5 | 11.34/5.02 | 19.17/0.25 | 106.13/0.02 | 0.74/0.02 |
| 6 | 14.66/5.96 | 22.32/0.32 | 191.06/0.02 | 0.96/0.02 |
| 7 | 14.23/6.24 | 24.45/0.38 | 176.92/0.02 | 1.15/0.02 |
| 8 | 13.42/5.97 | 21.17/0.38 | 169.25/0.02 | 1.88/0.02 |
| 9 | 13.05/5.99 | 21.29/0.39 | 157.95/0.01 | 4.17/0.02 |

| $n$ | Machine Learning Algorithms | | | |
|---|---|---|---|---|
| | *NB* | *SVM* | *PART* | *RF* |
| 10 | 13.36/5.95 | 27.26/0.36 | 159.1/0.01 | 6.07/0.02 |

\* training / prediction overhead

## V. DISCUSSION

We evaluated two different types of $n$-opcode information in our evaluation. One is the binary $n$-opcodes and the other is the frequency $n$-opcodes. The binary $n$-opcodes only describe what $n$-opcodes have been used in an application while the frequency $n$-opcodes also contains how many times each $n$-opcode has been used. From our evaluation, we observed that the binary $n$-opcodes more accurate than the frequency $n$-opcodes. Another advantage of the binary $n$-opcode is that we can reduce the storage overhead as mentioned in [9].

However, the frequency $n$-opcodes show very good accuracy when $n$ is small. This means that the frequency $n$-opcodes with small $n$ can be chosen for light-weight use case scenarios. Because the frequency $n$-opcodes tell us how many times each $n$-opcode has been used so this information can help to distinguish applications that use the same $n$-opcodes but in different numbers. The combination of the binary and frequency $n$-opcodes remains as future work.

## VI. CONCLUSION

In this paper, we investigated and analyzed $n$-opcode based static analysis approach to Android malware detection. This approach eliminates the need for hand-crafted or defined which require expert knowledge or analysis. Unlike most previous works that utilize defined features like API calls, permissions, intents and other application properties, our method allows for automatic extraction and learning of features from given datasets. Furthermore, we achieved analysis with longer $n$-grams than the state-of-the art by utilizing up to 10-opcodes in our experiments compared to the currently reported maximum of 5 in the literature. This was possible using a data segmentation technique during pre-processing in order to enable feature selection on as large a dataset as possible. Our results showed that by using frequency $n$-opcodes with low $n$, good classification accuracy can be achieved. Nevertheless, a maximum f-measure of 98% in both malware classification and categorization were obtained with $n=3$ and $n=4$, respectively.